\documentclass[page-classic]{epl2} 

\pdfoutput=1
\usepackage[english]{babel}
\usepackage{graphicx}

\title{The network organisation of consumer complaints}

\author{Luis Enrique Correa Rocha\inst{1}\thanks{E-mail: \email{Luis.Rocha@tp.umu.se}} \and Petter Holme\inst{1,2}}
\shortauthor{Luis E. C. Rocha \etal}

\institute{                    
  \inst{1} IceLab, Department of Physics, Ume{\aa} University, 90187 Ume{\aa}, Sweden\\
  \inst{2} Department of Energy Science, Sungkyunkwan University, Suwon 440--746 Korea}

\abstract{Interaction between consumers and companies can create conflict. When a consensus is unreachable there are legal authorities to resolve the case. This letter is a study of data from the Brazilian Department of Justice from which we build a bipartite network of categories of complaints linked to the companies receiving those complaints. We find the complaint categories organised in an hierarchical way where companies only get complaints of lower degree if they already got complaints of higher degree. The fraction of resolved complaints for a company appears to be nearly independent on the equity of the company but is positively correlated with the total number of complaints received. We construct feature vectors based on the edge-weight -- the weight of an edge represents the times complaints of a category have been filed against that company -- and use these vectors to study the similarity between the categories of complaints. From this analysis, we obtain trees mapping the hierarchical organisation of the complaints. We also apply principal component analysis to the set of feature vectors concluding that a reduction of the dimensionality of these from $8827$ to $27$ gives an optimal hierarchical representation.}

\begin{document}

\maketitle

Feedback in the form of complaints is both a way for improvement and an institutionalised legal right of consumers in many countries~\cite{Chandrashekaran07,Brewer03,Davidow03}. Complaining behaviour is a complex phenomenon not yet fully understood. It depends on several factors such as the cultural environment, gender, age and social status, the type of service or product, previous knowledge and even the establishment location~\cite{Nelson70,Kolodinsky95,Johnson02}. It is argued, for instance, that a primary cause of both off- and online complaints is unmet consumer expectations~\cite{Cho02}. Complaining can be communicated in different ways---directly (by contacting the companies) or indirectly (to other consumers)~\cite{Chelminski01}. Researchers in the social sciences have been studying how these variables affect consumer's behaviour and their propensity to complaining (or refrain from it) and how to improve consumer's satisfaction and loyalty~\cite{Nelson70,Kolodinsky95,Cho02,Crie03}.

Many countries have specific laws regarding rights and restrictions of consumers and companies. When a consensus between the parts is unreachable, the final decision is made by the legal authorities~\cite{ConsumerBR}. In this letter, we study the relation between filed complaints about different public and private organisations (henceforth simply referred to as \textit{companies} though they also include public organisations like schools and register offices) in Brazil. This system is large enough that network modelling can be useful to characterise its global organisation. Related datasets where network approaches have proved fruitful include trade patterns, trust and infrastructure webs, and online interaction systems~\cite{Hidalgo07,Huang07,Shang09}. We map the system into a bipartite network by letting companies and categories of complaints (henceforth we will talk of the categories simply as \textit{complaints}) represent two types of vertices and connect companies and complaints according to the data. Networks that, like in our case, connect two types of vertices are called \textit{two-mode networks} and form a special class of network models incorporating more information at the price of that many analysis methods for simple graphs are inapplicable~\cite{Li04,Holme07}.

We obtain our information about complaints and companies from the website of the Brazilian Department of Justice (in the section ``Consumers Rights''). The dataset is public and available as an electronic file. The file contains a list of public and private companies; each item has a set of categories of complaints (henceforth we will talk of the categories simply as \textit{complaints}) reported by consumers. One can also see if the complaint is resolved or not and to which one of six \textit{classes} it belongs to. There are $6$ classes of complaints and they are defined by the authorities according to the subject of the complaints. In the ``products'' class, for example, one (category of) complaint reads ``the delivered product is different from the order'' or in the class ``health'': ``[the company is] refusing to reimburse medical expenses covered by private insurance''. 

Some companies appear as different entries in the data. This might happen for several reasons, for instance the company has branches in different cities or is known by different names. Fortunately, a large number of them have an official register number (which sometimes is listed). We use this number, when available, to correct for the multiple entries. The database covers $19$ of the $27$ Brazilian states (containing $59.2\%$ of the population) between September $2007$ and August $2008$.

Company complaints data can be represented as a bipartite network $\Gamma_{\mathrm{bi}}$, defined by assigning vertices of different types to complaints and companies. If a complaint is made about a company, then the company and the complaint are connected by an edge. The number of similar complaints about the same company defines an edge-weight. The edges are split into two sets corresponding to solved and unsolved complaints, respectively.

\begin{figure}
  \onefigure[width=0.49\textwidth]{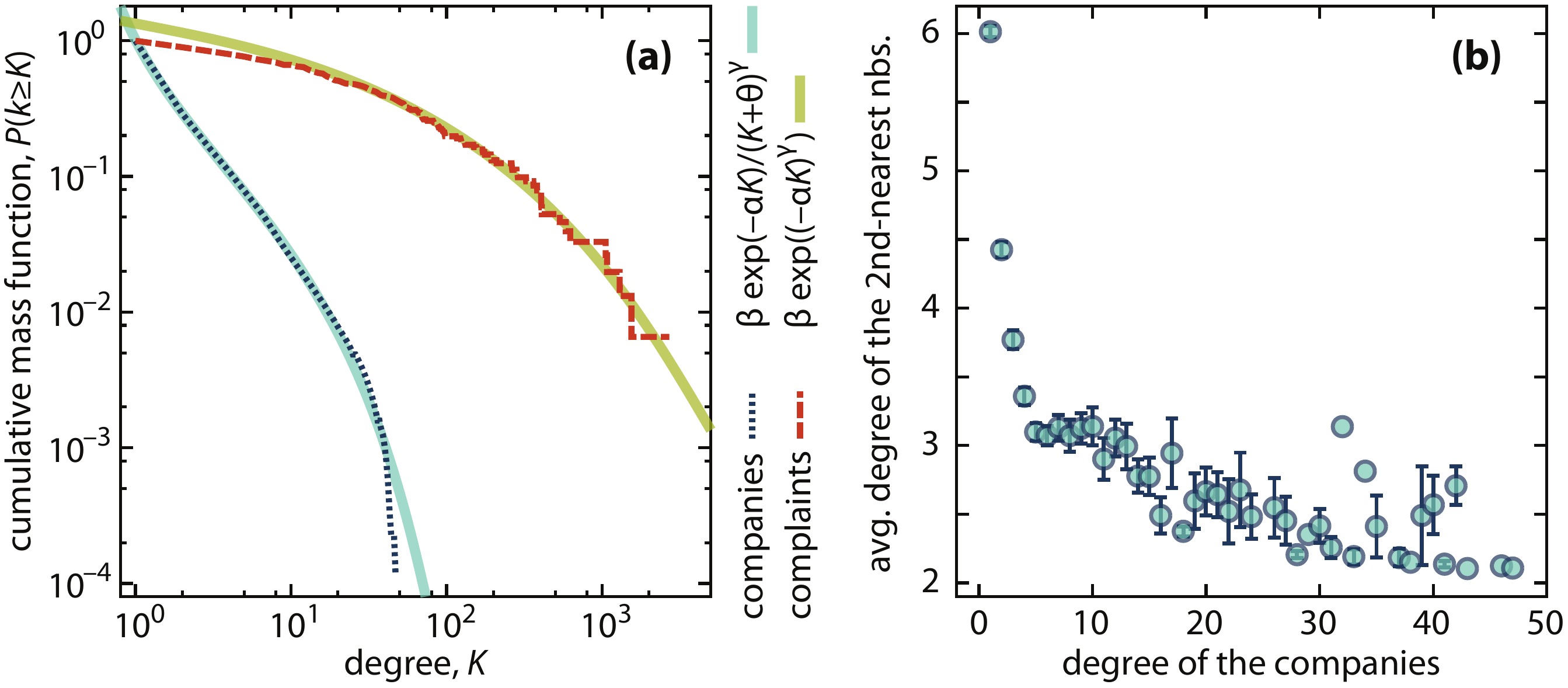}
\caption{(a) Cumulative degree distributions for the complaints and companies and least-squares fitting by using the functions $\beta \exp(-(\alpha K)^\gamma)$ where, $\beta=2.6\pm0.2$, $\alpha=0.22\pm0.04$, and $\gamma=0.289\pm0.006$ (complaints); and $\beta \exp(-\alpha K)/(K+\theta)^{\gamma}$ where, $\beta=0.489\pm0.006$, $\alpha=0.058\pm0.003$, $\theta=0.53\pm0.01$, and $\gamma=1.04\pm0.02$ (companies). The number after ``$\pm$'' stands for standard error. (b) Average degree of second neighbours of a reference company-vertex with degree $k$ as a function of $k$.}  \label{fig:01}
\end{figure}

The network contains many more companies than complaints (see table~\ref{tab:01}) and the average degree (number of neighbours) $k$ is higher for complaints than the companies (table~\ref{tab:01}). The cumulative degree distributions are also different for the two types of vertices. The curve for the complaints are reasonable fitted by a stretched exponential function $P(k\geq K)=\beta \exp(-(\alpha K)^\gamma)$, while the corresponding curve for companies fits well to a power-law with exponential cut-off $P(k\geq K)=\beta \exp(-\alpha K)/(K+\theta)^{\gamma}$ (figure~\ref{fig:01}-a). Stretched exponentials can emerge from sublinear preferential attachment~\cite{Krapivsky00}, but we will not speculate more about this. To estimate local correlations in the structure, we compare the actual network with a randomised version, where the degree distributions and bipartivity are conserved but the rest is randomised. The diameter (the longest distance --- length of the shortest-path --- between any pair of vertices) is the same after randomisation, and the average distance is slightly smaller than in the random version. The number of $4$-cycles (closed paths of length $4$, which measures the local redundancy of edges) is also slightly smaller in the original network (table~\ref{tab:01}). In sum, the network structure, in the unweighted network representation, is in many respects close to what would be expected from a random null model.

\begin{table}
\caption{Network measures for the original and the randomised version of the complaints and companies network (see explanation in the text). The abbreviation s.e.\ stands for standard error.} \label{tab:01}
\begin{center}
  \begin{tabular}{  c  c  c }
\hline
                 & Companies       &  Complaints              \\ \hline
No.\ vertices    & $8827$          &  $152$                   \\
Avg.\ degree     &    $2.04$       &  $118.28$                \\ \hline
                 & Original        & Random (s.e.)            \\ \hline
Avg.\ distance   &    $3.675$      &    $3.750$ ($0.003$)     \\
Diameter         &    $8$          &    $8$ ($0$)             \\
$4$-cycles ($\times10^6$)&  $2.40$ &    $2.69$ ($0.05$)       \\
Assortativity    &   $-0.283$      &    $0.000$ ($0.005$)     \\ \hline
  \end{tabular}
\end{center}
\end{table}

There is however one aspect where the unweighted network shows correlations --- the correlations between the degrees of nodes at either side of an edge. A straightforward measure of such correlations is the assortativity defined according to equation~\ref{eq:01}, where $k_C$ and $k_S$ correspond to the degrees of complaints and companies, respectively. The network is effectively disassortative (table~\ref{tab:01}). The disassortativity means that high-degree complaints have a tendency to be connected to low-degree companies and similarly, low-degree complaints to high-degree companies. In other words, rare complaints (low-degree complaints, i.e.\ those that are not associated to many companies) are more likely to be reported about companies that receive a broad range of complaints. Likewise, companies with few complaints usually receive the most frequent complaints. If a company receives a complaint, it will most likely be a common one, however, as long as the company continues getting complaints, they tend to get more specific. 
\begin{equation}
\label{eq:01}
  r = \frac{\langle k_Ck_S \rangle - \langle k_C \rangle \langle k_S \rangle}{\sqrt{\langle k_C^{2} \rangle - \langle k_C \rangle^2}\sqrt{\langle k_S^{2} \rangle - \langle k_S \rangle^2}}
\end{equation}
If we measure how the average degree of the neighbours at distance $2$ of a reference vertex $i$ is related to its own degree, we observe a small negative dependency (figure~\ref{fig:01}-b). Companies receiving diverse complaints tend to be connected to companies receiving few complaints. Notice that these neighbours two steps away are those vertices of the same type (representing the companies) that are connected to the same complaint as the reference vertex $i$ with degree $k_i$ (also representing a company)~\footnote{These are the same neighbours as those on the projected one-mode network of companies---where companies are connected if they have a common complaint---however, the degree in this case is not the same as the degree in the projected network.}. This suggests that highly reported companies (large degree) tend to belong to different sectors and not form clubs.

\begin{figure}
  \onefigure[width=1\linewidth]{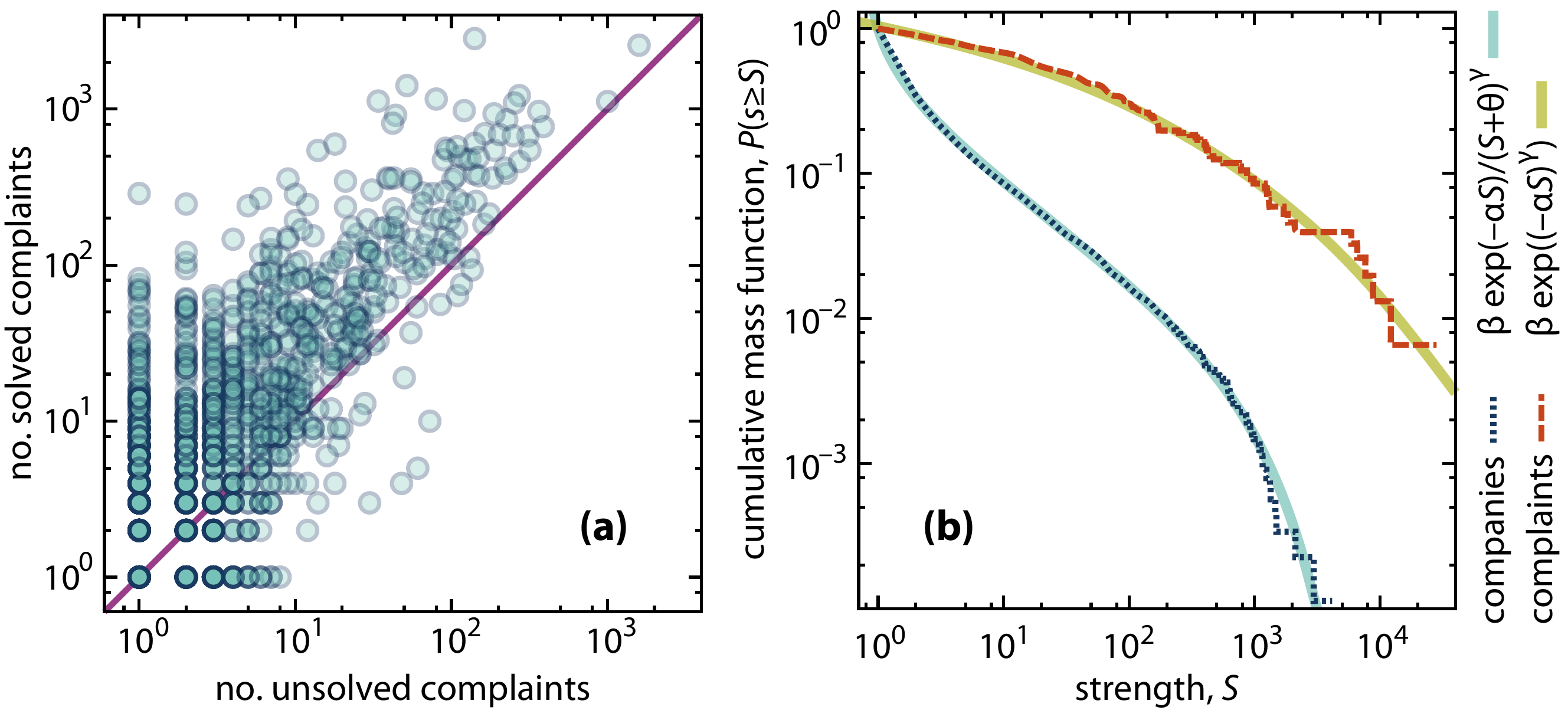}
  \caption{(a) Correlation between the number of solved and unsolved complaints. (b) Cumulative strength distributions for the complaints and companies, and the trends illustrated by least-squares fits. The complaints are fitted to $\beta \exp(-(\alpha S)^\gamma)$ where, $\beta=2.7\pm0.2$, $\alpha=0.7\pm0.1$, and $\gamma=0.187\pm0.002$; the companies by $\beta \exp(-\alpha S)/(S+\theta)^{\gamma}$ where, $\beta=0.3855\pm0.0004$, $\alpha=(9.27\pm0.06)10^{-4}$, $\theta=0.7630\pm0.0006$, and $\gamma=0.6629\pm0.0004$.}  \label{fig:02}
\end{figure}

Unweighted measures discard all information about the number of complaints against a company. This extra information is important because it shows that for the same company, some complaints are more likely to be reported than others. The total number of complaints $C_T$ of each type is used as weights to the edges in the bipartite network. We separate solved and unsolved complaints. There is a positive correlation (Pearson's correlation coefficient is $0.72$) between the number of solved and unsolved complaints for a company (with typically more solved than unsolved complaints), see figure~\ref{fig:02}-a. If we sum all the weights of the edges connected to a vertex $i$, we obtain the \textit{vertex strength} $s_i$. Considering the total number of complaints $C_T$, the cumulative strength distributions in figure~\ref{fig:02}-b follow the same functional forms as for the degree distributions. In the case of companies, the exponential cut-off is smaller for the strength in comparison to the degree. On the other hand, $\theta$ is smaller for the degree distribution, suggesting that the number of complaints about a company does not vary linearly with the total number of reported complaints.

\begin{figure}
  \onefigure[width=0.46\linewidth]{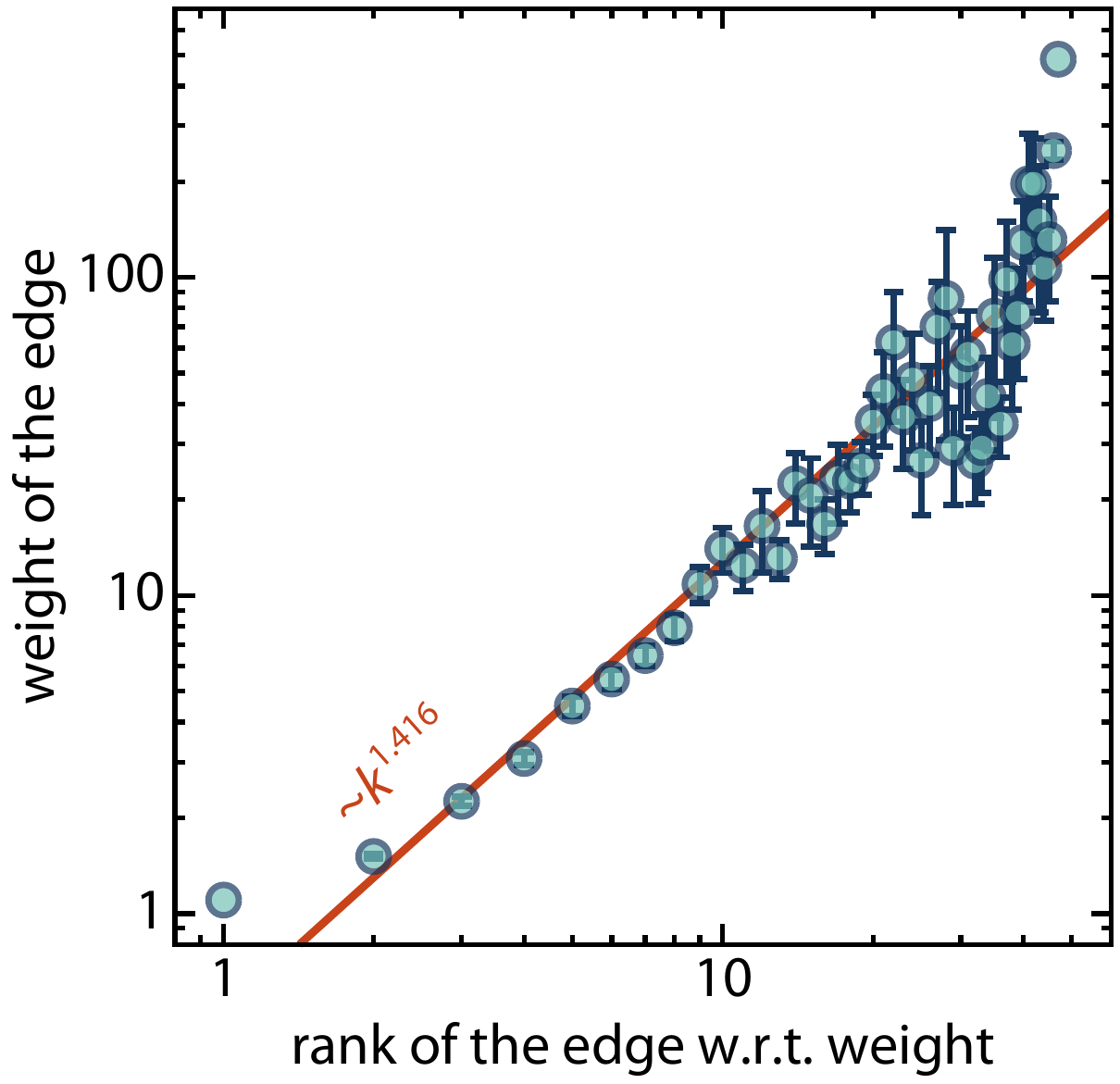}
  \caption{(a) Total number of complaints, or weight, of an edge as a function of the rank of the edge ranked with respect to (w.r.t.)\ the weight. The average value and the respective standard error are taken over all companies for each rank value.} \label{fig:03}
\end{figure}

The dissassortativity observed in the previous section indicate that, on average, low-degree companies tend to receive high-degree complaints and, as long as the degree of the company increases, it tends to connect to lower degree (more specific) complaints. If we rank the edges of each company according to their weights, figure~\ref{fig:03} shows that averaging over all companies, the total number of complaints increases super-linearly with the rank. It means that, on average, a company is reported more times about previously reported complaints than about new ones.

\begin{figure}
  \onefigure[width=0.97\linewidth]{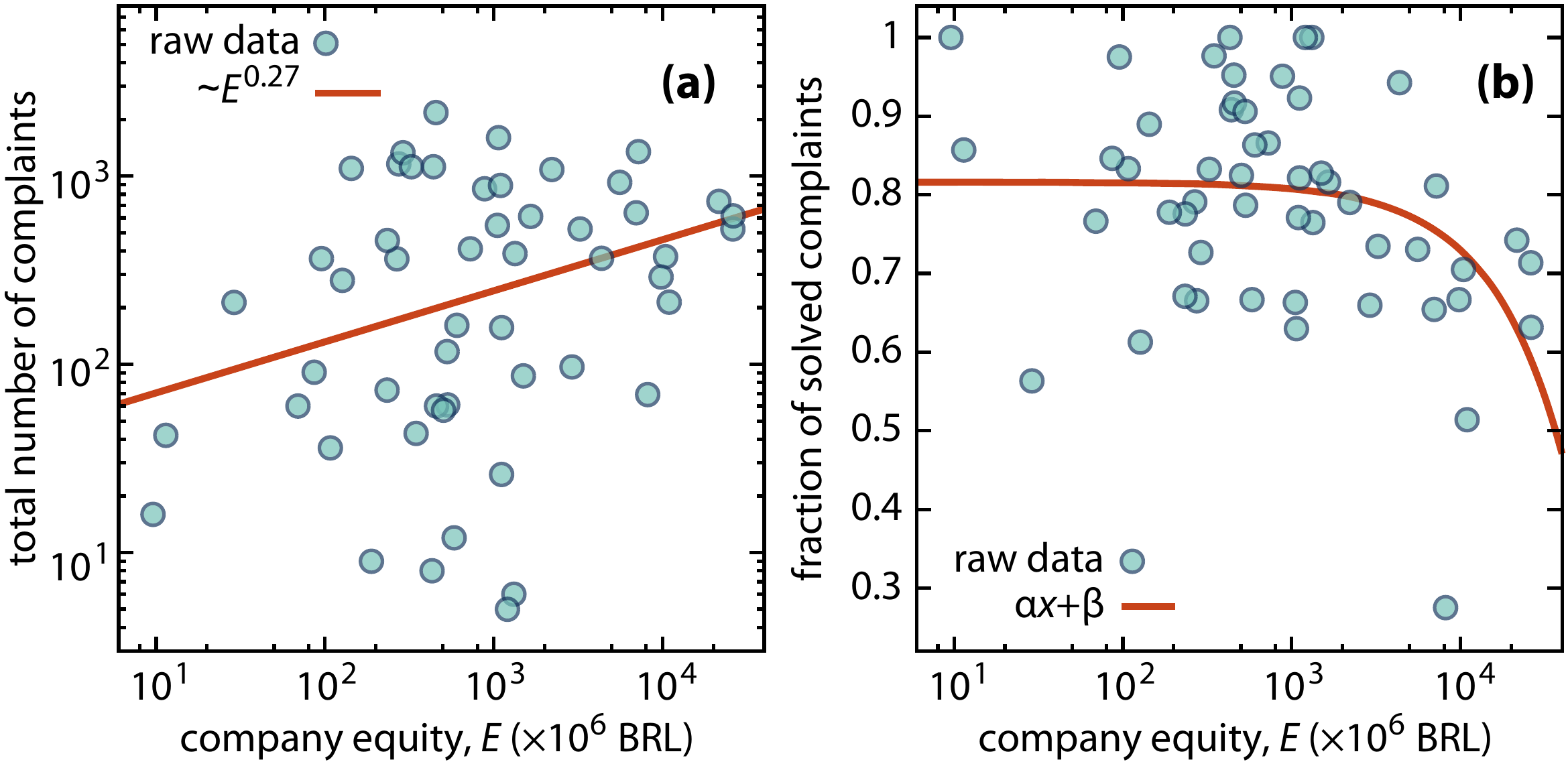}
  \caption{(a) Total number of complaints versus the equity value of the company (and least-square fits to show trends). (b) Percentage of solved complaints versus the equity value of the company. The Pearson correlation coefficient is $-0.37$ and the linear least-square fitting, $\alpha x+\beta$, gives $\alpha=(-9\pm3)10^{-6}$ and $\beta=0.82\pm0.02$. The abbreviation BRL means Brazilian Real. The abscissa is logarithmic.}  \label{fig:04}
\end{figure}

The number of complaints increases sublinearly with the size of the company (figure~\ref{fig:04}-a), here measured by the equity---essentially the difference between the assets and the liability---and the fraction of solved complaints is nearly independent of the equity (figure~\ref{fig:04}-b). These results suggest that even though companies solve more complaints the more complaints they get, this is not directly connected to their wealth (figure~\ref{fig:04}-b). This can be explained if the consumer's support departments are simply driven by consumers' needs and requests, and not limited by the size of the company.

The edge-weights provide more information about the relations between the two types of vertices. To identify the structural similarity between different vertices incorporating this information, we create a vector containing both the local connectivity of a vertex and the weights of the respective edges~\cite{Costa2hier,Costa3trees}. With this \textit{feature vector}, we quantify the topological similarity between the vertices and compare with known classes from the dataset. The patterns of connections in a network can reflect intrinsic properties of the vertices. For example, the company's sector restricts the possible complaints it can receive. Consequently, the pattern of connections of one vertex can be used to create a topological identity~\cite{Costa2hier,Costa3trees} to be further related to known properties of the vertex. A simple approach to create a feature vector is to use the number of neighbours and the weights of the respective edges as the features of this vector. Though this methodology could be applied to any set of vertices, we only create one vector $\mathbf{v}_i$ for each different complaint, where the entry $k$ contains the total number of complaints $C_T$ of type (or category) $i$ about company $k$. With our network, this procedure creates $152$ vectors in an $8827$-dimensional space~\footnote{Note that the same methodology could be applied to the companies, i.e.\ each company has one feature vector and each entry of this vector corresponds to a complaint.}.

To quantify the similarity between two vertices with respect to their feature-vectors, we calculate the cosine similarity between the respective two vectors $\mathbf{v}_i$ and $\mathbf{v}_j$ (equation~\ref{eq:02}), where $|\;\cdot\;|$ is the magnitude of the vector. Among the different similarity measurements, the cosine similarity captures the trend to the cost of disregarding the magnitude. A consequence is a similarity scale ranging from $0$ (least similar) to $1$ (most similar).
\begin{equation}
\label{eq:02}
  \mbox{cosine similarity} = \frac{\mathbf{v}_i \cdot \mathbf{v}_j}{| \mathbf{v}_i |\; | \mathbf{v}_j |} .
\end{equation}

\begin{figure*}
  \onefigure[width=0.66\linewidth]{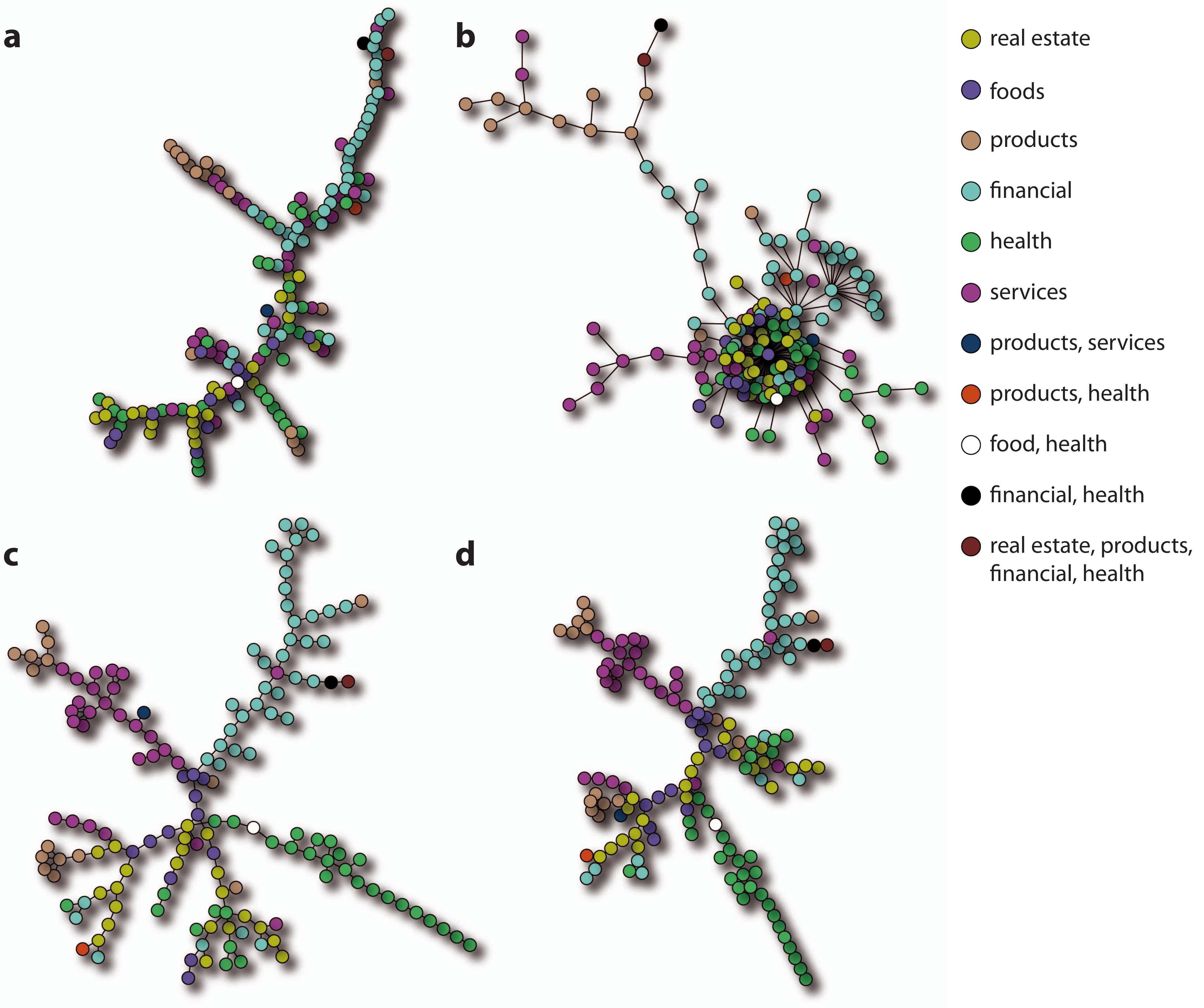}
  \caption{Minimum spanning tree for different number of eigenvalues. Considering only the PCs corresponding to (a) the four largest eigenvalues; (b) all PCs, i.e.\ $8827$ dimensions; and the optimal numbers (c) $27$ and (d) $28$ eigenvalues. The different colours represent different groups of complaints as formed by the $6$ different classes available in the data: real estate, food, products, financial, health and services. }  \label{fig:06}
\end{figure*}

The structural similarity between all pairs of vertices is mapped into a new fully connected, weighted network $\Gamma_{\mathrm{sim}}$, where each vertex corresponds to a different complaint and the edge-weights correspond to the cosine similarity between them~\cite{Costa3trees}. A structure connecting similar vertices is obtained by using the minimum spanning tree (MST)~\cite{graph,Tumminellomst07}. This algorithm creates a tree $\Gamma_{\mathrm{MST}}$ containing all original vertices (i.e.\ the complaints) and minimises the total weight on the edges or as in our case, it maximises the weight since higher values correspond to higher similarity. Consequently, similar vertices are maintained connected in an optimal tree-structure such that the sum of the weights is maximum (figure~\ref{fig:06}-a).

\begin{figure}
  \onefigure[width=1\linewidth]{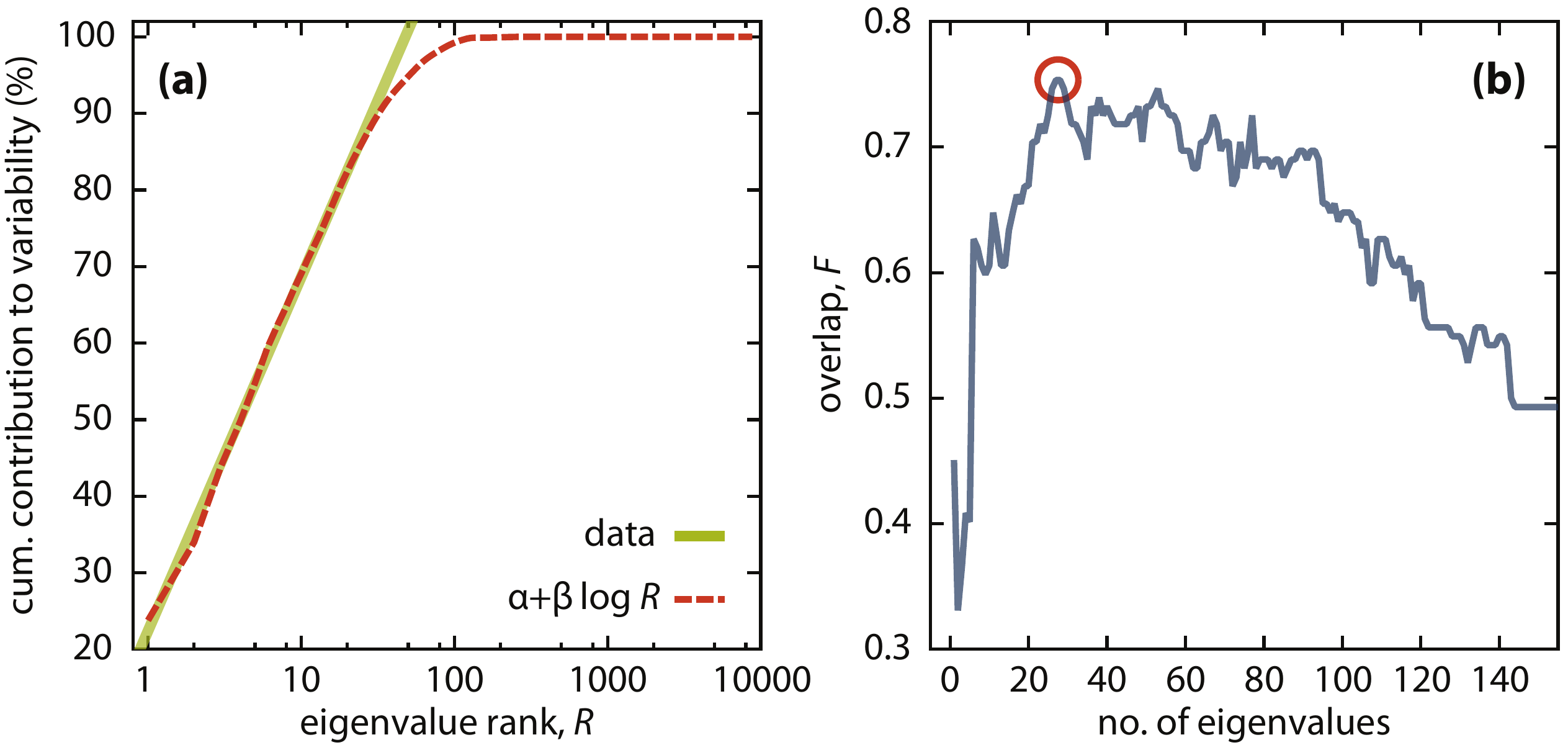}
  \caption{(a) Contribution of the eigenvalues for the total variability of the data. The eigenvalues $\lambda$ are ranked on increasing values, $R$. A logarithmic function is fitted for small ranked eigenvalues, $f(R) = \alpha + \beta\log(R)$. (b) Fraction of correctly identified pairs of complaints of the same class as a function of the number of eigenvalues. The peak is centred around $27$ and $28$ eigenvalues. The eigenvalues are chosen in increasing value of contribution to variability according to the PCA method.}  \label{fig:05}
\end{figure}

The feature vectors can be very correlated. If some dimensions are correlated we can assume they are explained by the same underlying mechanisms and thus, by merging them, simplify the representation of the system without loosing much information. The multivariate method \textit{Principal component analysis} (PCA) is a general method for this type of decorrelation and data reduction~\cite{Jolliffe02,Costello05}. In the PCA, the axes in the original $n$-dimensional space are rotated to point towards the maximum variability of data. The rotation matrix is the covariance matrix of the different dimensions and the eigenvalues of this matrix provide a scale of data variability in the directions corresponding to the eigenvectors --- \textit{principal components} (PCs) --- of the rotation matrix. Applying this rotation to our data, we obtain the original set of points projected into the new $n$-dimensional space~\footnote{This method is also known as singular value decomposition.}. Therefore, PCs corresponding to small variability (small eigenvalues) and consequently, little associated information, can be discarded without loss of relevant information. By using the projected points, we construct new feature vectors, for all complaints, using solely selected variables. We, then, repeat the procedure of calculating the new similarity between any two vectors and assign the value to the weights in the network of complaints. Finally, we perform the MST procedure again and obtain a new tree connecting similar complaints. These trees are expected to be different if we select different PCs to form the feature vectors. Before performing the PCA, the datapoints are replaced by their Z-scores (normalisation) such that all dimensions have zero mean and standard deviation equal to one. This procedure highlights the variability contribution of the new axes after performing a PCA rotation. By doing this standardisation, the sum of all eigenvalues is equal to the number of points, since this corresponds to maximum variability. Furthermore, all eigenvalues larger than one provide more information than any of the original axes. Since there is no rule to choose the optimal eigenvalues, one procedure is to count the contribution of each eigenvalue in the total variability by dividing its value by the sum of all eigenvalues. In our data, the contribution of the eigenvalues follows a logarithmic function for the most relevant eigenvalues if ranked on increased values (figure~\ref{fig:05}-a). We see a variability saturation starting at about $30$ eigenvalues, where increase in the number of eigenvalues brings small contribution in the variability. Essentially, almost all variability can be described by using less than $100$ of the largest eigenvalues. More specifically, the $136$ eigenvalues larger than one account for $99.97\%$ of the variability.

Previously, we found that a vast majority of the dimensions of the feature vectors can be removed without losing relevant information. Since the complaints are classified into different classes in the database, we use this information to search for the optimal number of eigenvalues needed to cluster similar vertices. A direct measure of clustering is to count the fraction $F$ of edges connecting vertices of the same group as defined into the dataset (eqn~\ref{eq:03}). There are $6$ classes but we identify $11$ groups, since some complaints are simultaneously associated to more than one class.

\begin{equation}
\label{eq:03}
  F = \frac{\mbox{no.\ edges between vertices same group}}{\mbox{no.\ edges in the tree}-(\mbox{no.\ groups} - 1)}
\end{equation}

Figure~\ref{fig:05}-b shows the dependence of the fraction $F$ with the number of eigenvalues used (choosing them in increasing value). We observe a peak at $27$ and $28$ eigenvalues corresponding, respectively, to $87.99\%$ and $88.56\%$ of the variability. By using all eigenvalues (see figure~\ref{fig:06}-b), the matching gets better than with few eigenvalues (compare to figure~\ref{fig:06}-a) but still, it is worse than for the optimal value (figures~\ref{fig:06}-c,d). From figure~\ref{fig:06}-b we also see that the majority of the vertices are connected to a single central vertex and do not show a clustered structure. In contrast, at the optimal number of eigenvalues, multiple branches emerge where some are essentially formed by vertices in the same class, as for instance, those related to financial, health and services (figure~\ref{fig:06}-c,d). The vertices corresponding to real estate and food related complaints are clustered more centrally in the tree. Interpreting the tree needs some degree of caution. The MST algorithm gives an optimal tree structure such that the total weight is minimised (maximised in our case) and each vertex is connected to at least one other vertex. Note that two vertices might be more similar than those connected in the tree. Considering one pair, the similarity is never larger than the largest in the path between two vertices in the MST.

To summarize this letter, we study the relation between different public and private companies according to the complaints they have obtained. We map these relations between companies and complaints to a bipartite network and see that degree and strength distributions are reasonable well described by strechted exponentials for complaints, while power-laws with exponential cutoff are fitted to the company-vertices. If considering the companies as passive objects that can be selected by individuals (via a specific complaint), the system becomes close related to web-based user-object systems, where a user selects objects (e.g.\ movies) according to its personal preferences. Similar to our results, Shang and collaborators found that object-degree distribution is better described by a power-law functional form while for the user-degree distribution, an strechted exponential is more adequate~\cite{Shang09}. The network has small dissassortativity, which means that companies with many different complaints have a tendency to be connected to less common complaints. Consequently, common complaints are usually connected to companies that receive few complaints. The results suggest that initially, a company receives common complaints and, as long as the number of complaints increases, they tend to be more specific. Related to this effect, we observe a superlinear relation in the number of reported complaints where companies tend to proportionally receive more complaints previously reported than new ones. The first effect is observed in user-object systems as well, i.e.\ popular objects tend to be selected by new users while very active users preferably select those less commonly chosen objects~\cite{Shang09}. These properties have been applied to enhance accuracy of personal recommendation systems~\cite{Liu10} and might be used in the context of management to better identify deficiencies in the company organisation, products or services.

By using the local network structure and corresponding edge-weights, we create a multidimensional vector of topological features for each complaint-vertex. From this vector, we calculate the similarity between different vertices, and by using the minimum spanning tree algorithm, we extract a tree connecting the most similar vertices. To reduce the correlation and noise in the data, we perform the principal component analysis and compare the clustering in case of a different number of dimensions in the rotated datapoints. The results show significant differences according to the amount of information we include in the vector of features, i.e.\ more or less dimensions. By comparing with annotated classes from the database, we identify that only $27$ or $28$ dimensions, $0.32\%$ of the original number, are needed to provide $88\%$ matching. This indicates that there are strong correlations in the data, not captured by other network measures. Some classes of complaints are reasonable well clustered in branches of the resulting tree, especially those related to financial, health and services sectors. On the other hand, complaints related to real estate and food sectors were identified more centrally in the tree. It is an open question for the future whether our observations are specific of our data or universal across societies. 

\acknowledgments
The authors are grateful to Aaron Clauset and Tam\'as Nepusz for comments, and to the Swedish Foundation for Strategic Research, the Swedish Research Council and the WCU program through NRF Korea funded by MEST (R31--2008--000--10029--0) for financial support.

\end{document}